\documentclass[pre,aps,floats,superscriptaddress,floatfix,twocolumn]{revtex4}
\usepackage{amssymb,amsmath}
\usepackage{amsmath,amssymb}
\usepackage{graphicx}
\usepackage{psfrag}
\usepackage{color}
\usepackage{soul}
\usepackage{dcolumn}
\usepackage{bm}
\usepackage[normalem]{ulem}

\def\beq{\begin{equation}}
\def\eeq{\end{equation}}
\def\bea{\begin{eqnarray}}
\def\eea{\end{eqnarray}}
\begin{document}
\title{
Defect versus defect: stationary states of single file marching in periodic landscapes with road blocks
}
\author{Atri Goswami}\email{goswami.atri@gmail.com}
\affiliation{Gurudas College, 1/1, Suren Sarkar Road, Jewish Graveyard,
Phool Bagan, Narkeldanga, Kolkata 700054, West Bengal, India}
\affiliation{Barasat Government College,
10, KNC Road, Gupta Colony, Barasat, Kolkata 700124,
West Bengal, India}
\author{Rohn Chatterjee}\email{rohn.ch@gmail.com}
\author{Sudip Mukherjee}\email{sudip.bat@gmail.com,aca.sudip@gmail.com}
\affiliation{Barasat Government College,
10, KNC Road, Gupta Colony, Barasat, Kolkata 700124,
West Bengal, India}

\date{\today}

\begin{abstract}
 Totally asymmetric simple exclusion process (TASEP) sets the paradigm for one-dimensional driven single file motion. We study a periodic TASEP with two ``road blocks'' or defects of different kinds, one point and another extended, across which particle flows are inhibited. We show how the interplay between particle number conservation and competition between the defects lead to inhomogeneous steady states with localised domain walls (LDW). The LDW locations jump discontinuously, indicating a  { discontinuous} transition between these LDW states, as the system passes from being controlled by one defect to the other. When the defects are ``competing'', instead of an LDW a pair of delocalised domain walls   appear, none of which can penetrate the extended defect. A minimum current principle can be used to identify the dominant defect that controls the domain wall formations.   { Our results should be important in diverse systems, ranging from protein synthesis by ribosomes in biological cells to urban traffic networks.}
\end{abstract}

\maketitle

\section{Introduction}

Single file motion implies particle motion along quasi one-dimensional (1D)
narrow channels where the particles cannot cross each
other due to hardcore repulsion. This was originally
introduced by Hodgkin and Keynes~\cite{hodg} to describe ion
transport in biological channels. 
Unidirectional or driven single file motion have received increasing attention in the recent past for their
wide-ranging interdisciplinary applications. { We  study the role of bottlenecks and their interplay and competition in controlling the stationary states of driven single file motion, a question of potential relevance in wide-ranging systems, e.g., traffic flow ~\cite{ped1,ped2,ped3}, colloidal particles flowing through geometric constraints~\cite{coll1,coll2,coll3}  and ribosome translation along messenger RNA (mRNA) loops~\cite{mrna1}}. 

Totally asymmetric simple exclusion process (TASEP),  a paradigmatic example of 1D driven single file motion, consists of 1D lattice with $L$ sites,  with particles
 hopping  unidirectionally,  subject to exclusion. Originally proposed to describe protein synthesis by ribosome translocation along messenger RNA strands in eukaryotic cells~\cite{mac}, it was re-invented as a
 archetype for nonequilibrium phase transition in 1D
open systems~\cite{krug,tom}. { TASEP has been adopted to gain generic understanding about protein synthesis in cells~\cite{mrna1,tom-prl,beate1}. In particular, Refs.~\cite{tom-prl,beate1,krug-new} explored the sensitive dependence of the particle (``ribosome'') currents in an open TASEP with isolated or clusters of bottlenecks (``rare codons''), theoretical results that resonate with relevant experimental findings~\cite{bao-natcomm}; see also Ref.~\cite{mrna5}.  In a surprising cross-field connection, congestion in vehicular or pedestrian movements by road blocks - increasingly important in urban life, has also been addressed using TASEP-like cellular automata models with bottlenecks~\cite{comp}. While the effects of defects on the stationary currents in an open TASEP has already been studied, both in the context of protein synthesis in cells~~\cite{tom-prl,beate1,krug-new} and also from pure nonequilibrium physics standpoint~\cite{open1,open2,open3,atri1,sm-ab-tasep}, the question of identifying the dominant defect and its interplay  in a closed many-defect transport system remains open.} 

{ In this Article, we study a particle number conserving TASEP in a ring with one point and one extended  defects as a conceptual model for  competition among road blocks in a closed driven single file motion. We elucidate how the stationary density profiles of the TASEP emerge as an interplay between the dominant defect and number conservation.} Our principal results are (i) the defects lead to domain walls (DW) in the stationary density profiles connecting segments lower and higher than 1/2,
when the system  is neither nearly empty or nearly full, but rather has a moderate density. This defines a DW phase. {  In a striking display of defect competition, the DW location jumps {\em discontinuously} as the system undergoes a hitherto unstudied {\em first order} transition between point defect (PD) and extended defect (ED) controlled states.} (ii) In the DW phase, generally there is a single localised DW (LDW) for both PD or ED. Nonetheless, the LDW profiles for a PD and ED are fundamentally different from each other. (iii) In the event, the two defects are equi-dominant (in the sense explained below), instead of a single LDW, a pair of delocalised domain walls (DDW) emerges, forming  {\em outside} ED. The latter is in its maximal current (MC) phase. (iv) For a nearly empty or filled system, the density everywhere is less or more than 1/2, reminiscent of the low density (LD) or high density (HD) phases of a TASEP with open boundary conditions~\cite{krug,tom,rev1}. Our results on DWs can be reconciled within a minimum current principle.

 {  Our work should be useful for understanding  the phenomenologies of 
 ribosome translocations along
closed mRNA loops (circular translation of polysomal mRNA)  with clusters of slow
codons along which ribosome translocations are inhibited~\cite{mrna1,mrna11, mrna2,mrna3,mrna4,mrna5} and traffic jams or congestion in urban transport networks with multiple road blocks~\cite{ped2,ped3}. From a theoretical standpoint, our results complement existing research on the effects of a slow bond on 1D driven systems; see Refs.~\cite{soh,soh1}.}

\section{Model}
\begin{figure}[htb]
\includegraphics[height=4.1cm]{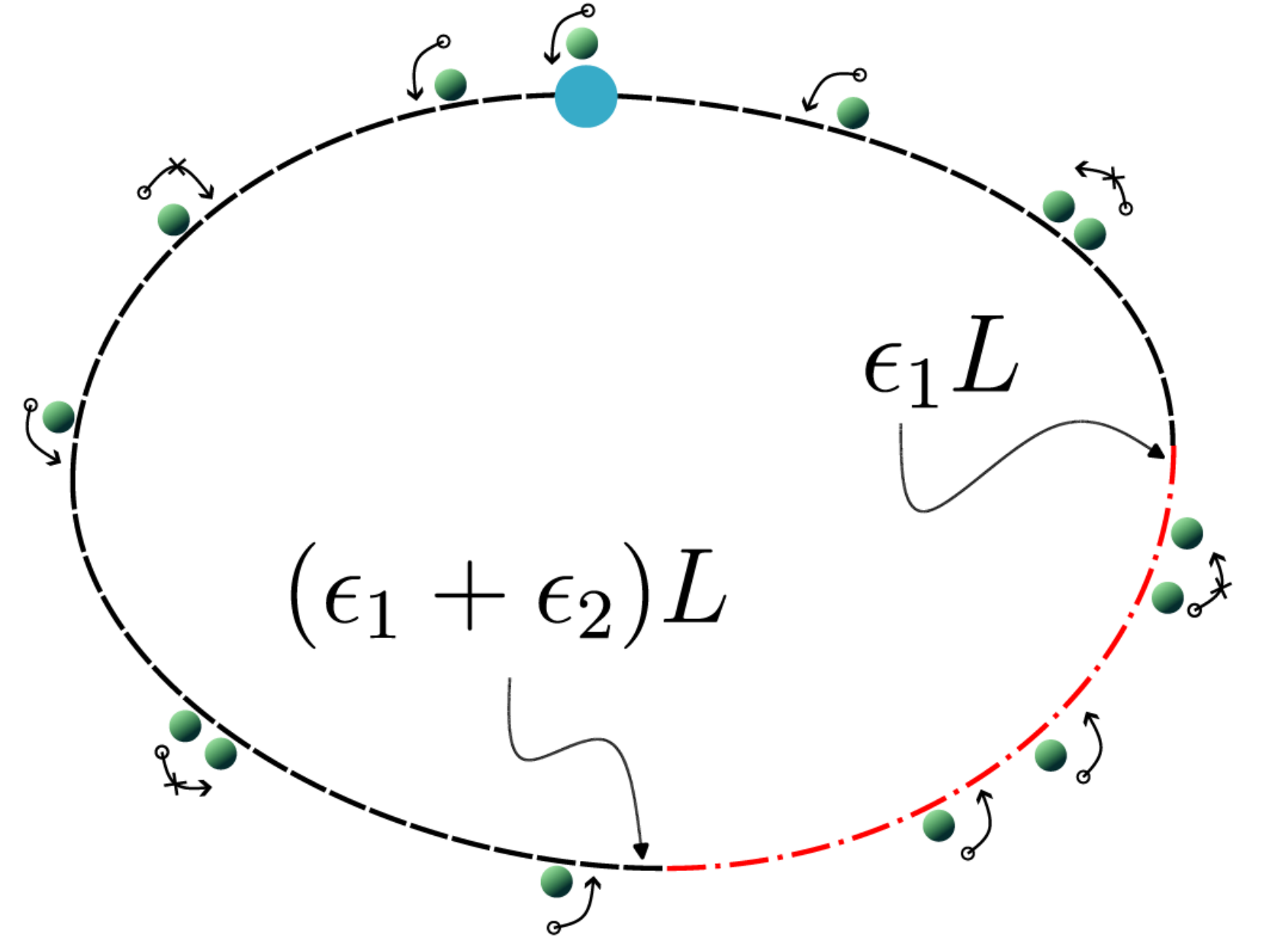}
 \caption{Schematic model diagram. The thick bluish green circle marks the point defect across which the hopping rate is $p<1$. The arc in red is the extended defect along which the hopping rate is $q<1$. In the remaining segments, the hopping rate is unity (see text).}\label{model-diag}
\end{figure}

\begin{figure}[htb]
 \includegraphics[width=7.8cm]{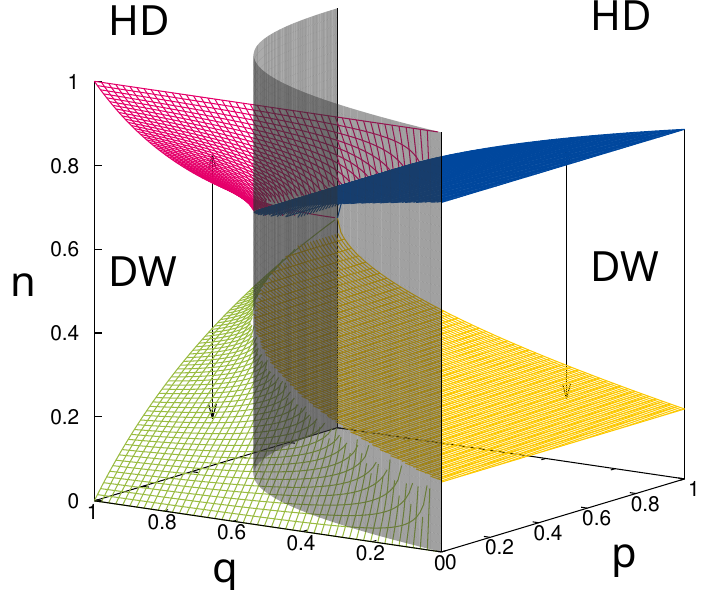}
 \caption{ Mean-field phase diagram of the model in the $p-q-n$ space showing the phases in the model. HD and DW (with one LDW) phases are marked. LD phase that lies below the DW phase in the figure is not visible. Points on the gray curved plane lying as a divider of the DW phase have $J_\text{PD}=J_\text{ED}$  satisfied, where $J_\text{PD},\,J_\text{ED}$ are the steady state currents in the PD and ED dominated situations respectively (see text below) and have a pair of DDWs fully or partially covering $T_1$ and $T_3$. In the DW phase region, on the left of this curved surface (i.e., with $J_\text{PD}<J_\text{ED}$), there is an LDW due to the point defect in either $T_1,T_2$ or $T_3$; on the right of this plane (i.e., with $J_\text{ED}<J_\text{PD}$), one has an LDW due to the extended defect in $T_1$ or $T_3$ and correspondingly an MC phase in $T_2$. In the HD (LD) phase region, one has HD-HD-HD (LD-LD-LD) phase in $T_1,T_2,T_3$ (see text).}\label{3d_phase_diagram}
\end{figure}

\begin{figure}[htb]
 \includegraphics[width=4.2cm]{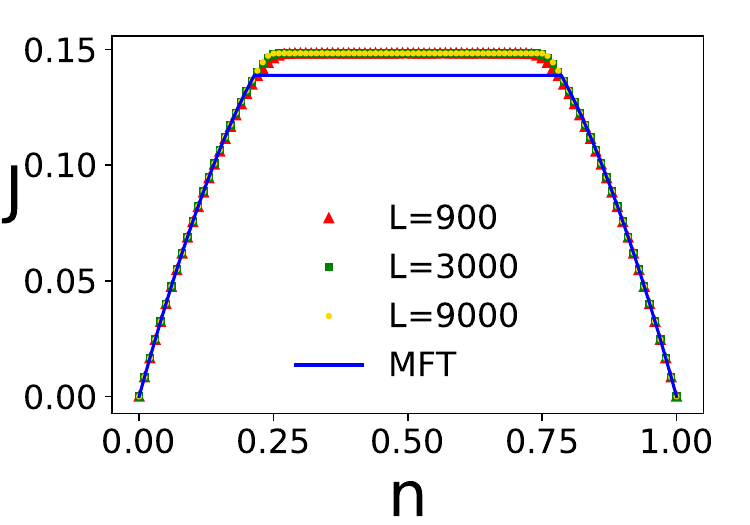}\hfill 
 \includegraphics[width=4.2cm]{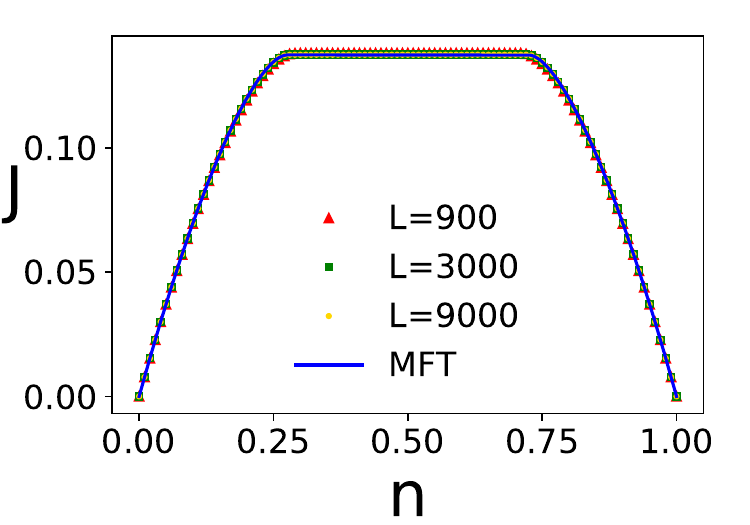}
 \caption{Fundamental diagrams ($\epsilon_1=\epsilon_2=1/3$) the phase space regions dominated by the {(left)} point defect ($p=0.2,q=0.65$), with visible mismatch between MFT and MCS results and some finite size effects, {(right)} extended defect ($p=0.2,q=0.55$), with good MFT-MCS agreement and no observable finite size effects (see text). }\label{funda}
\end{figure}

Our model, a periodic TASEP, has a single slow site (PD) at $i=1$ and an extended slow section (ED) from $i=\epsilon_1\, L$ to $(\epsilon_1+\epsilon_2)L,\,\epsilon_1,\epsilon_2<1$; see Fig.~\ref{model-diag} for a schematic diagram of the model. The hopping rate across PD is  $p<1$, and ED has a hopping rate $q<1$. In general $p\neq q$. Elsewhere in the system, the hopping rate is unity. The particles are assumed to move in the anticlockwise direction. It has a mean particle number density $n$, a constant of motion. 
The phases in the model and the associated transitions are controlled by $p,q,n$. The  mean-field phase diagram with $\epsilon_1=\epsilon_2=1/3$ is shown in Fig.~\ref{3d_phase_diagram}, giving the phases in the three segments - $T_1$ between $i=1$ to $\epsilon_1L$, $T_{2}$ between $\epsilon_1L$ to $(\epsilon_1+\epsilon_2)L$  and $T_{3}$ between $(\epsilon_1+\epsilon_2)L$ to $L$, connected serially. {   See Fig.~\ref{funda} {(left)} with $p=0.20, q=0.65$ and Fig.~\ref{funda} {(right)} with $p=0.20, q=0.55$ for the associated fundamental diagrams, giving the current-density relations, in the PD and ED dominated regions of the phase space, obtained from mean-field theory (MFT) and Monte-Carlo simulations (MCS). }

Thus, as $n$ rises from a very low value, the system moves from the LD-LD-LD phase to DW phases, dominated either by PD or ED, or both equally competing, for moderate $n$, eventually to the HD-HD-HD phase for very high $n$ close to unity. These qualitative descriptions are presented in  Movies 1, 2 ,3 in the Supplemental Material (SM)~\cite{sm}.

\section{Domain Walls}

In this section, we focus on the DWs. We derive the corresponding results on the DWs  within MFT~\cite{blythe}, supplemented by extensive MCS studies, which provide quantitative basis to the movies in SM.  The density profiles for the phases without any DWs, i.e., in LD-LD-LD and HD-HD-HD phases, can be obtained by using particle number and current conservations; see Appendix~\ref{all-ld-hd}. In addition, the absence of MC-MC-MC phase is argued in Appendix~\ref{all-ld-hd}.

To set uo our MFT, it is convenient to introduce a quasi-continuous coordinate $x\equiv i/L$ in the thermodynamic limit $L\rightarrow \infty$, with $0\leq x\leq 1$. Now define $\rho_a(x)\equiv \langle \rho^a_i\rangle$ as the local density, where $\langle...\rangle$ implies temporal averages in the steady states and $a=1,\,2$ or $3$ for the three segments. The stationary currents $J_1,\,J_2,\,J_3$ in the three segments in MFT are
\begin{eqnarray}
 J_1=\rho_1(1-\rho_1),J_{2}=q\rho_2(1-\rho_2),J_{3}= \rho_3(1-\rho_3)\label{curr1}
\end{eqnarray}
are all equal due to current conservation, where $\rho_1,\,\rho_2$ and $\rho_3$ are the stationary densities in $T_1,\,T_{2}$ and $T_{3}$ respectively. 
Particle number conservation gives
$ \int_0^{\epsilon_1}\rho_1 dx + \int_{\epsilon_1}^{\epsilon_1+\epsilon_2}\rho_2 dx +\int_{\epsilon_1+\epsilon_2}^1\rho_3dx = n$,
where $n=\sum_i\,n_i/L$ is the mean particle density in the system. Steady state currents (\ref{curr1}) together with particle number conservation can be used to calculate the steady state densities $\rho_1,\,\rho_2,\,\rho_3$. A complete characterisation of the steady states of this model requires specifying the phases in all of $T_1,\,T_2,\,T_3$, i.e., solving for all of $\rho_1,\,\rho_2,\,\rho_3$. While an open TASEP can be in LD, HD and MC phases, a TASEP in a closed system can also be in domain wall (DW) phase in an extended region of the control parameter space~\cite{lebo,niladri-tasep,mustansir,hinsch,tirtha-niladri,parna-anjan,astik-parna}. 
We will see below that  number and current conservations jointly ensure that {\em not all} of the possible phases are actually admissible in the present model.

The emergence of DWs, as revealed in our MCS results and observed in the movies~\cite{sm} can happen in two distinct ways.
Physically, from the LD-LD-LD phase upon addition of particles, $n$ rises 
{and it eventually reaches a lower threshold, a macroscopically nonuniform steady state in the shape of a DW is formed.
A DW, essentially a pile up of particles due to a bottleneck or a defect, should form {\em behind} the defect, point or extended, as the case may be, and then start to grow as $n$ rises further. While this remains true independent of the defect, our MCS results reveal a striking aspect - a DW that forms behind a PD can be anywhere in the TASEP ring covering all of $T_1,\,T_2,\,T_3$, being controlled by $n$, whereas a DW, which forms behind an ED, never enters $T_2$! This leads to complex density profiles when the two defects are ``competing''; see below and also Movie 3 in SM~\cite{sm}. To proceed systematically, setting aside for the time being the question of which of the two defects will have a DW behind it, let us separately consider DW formations by PD and ED. Considering a PD, a DW is first formed in $T_1$, as $n$ just exceeds $n_{cL}^p$,  a threshold for DW formation (see below) with the DW position $x_w$ being at $x=0$. As $n$ exceeds $n_{cL}^p$, the size of this pile or the DW increases by shifting its position $x_w$ (see below for a formal definition) along the ring, moving from $T_1$ first to $T_2$ and then to $T_3$, finally reaching $x=1$, at which point the DW ends at a threshold $n_{cU}^p$ of $n$ (see below), bringing the model to its HD-HD-HD phase. Now assume a DW at $x_w$ in $T_1$, i.e., $0<x_w<\epsilon_1$, connecting a high density segment with density $\rho_\text{HD}=1/(1+p)$ between $0$ and $x_w$ and a low density segment with density $\rho_\text{LD}=p/(1+p)$ between $x_w$ and $\epsilon_1$~\cite{lebo,niladri-tasep}. 
Then using number conservation
\begin{eqnarray}
 \rho_2&=&\left[\frac{1}{2}\pm\bigg\{\frac{1}{4}-\frac{p}{q(1+p)^2}\bigg\}^{1/2}\right]\nonumber \\
 &=& \rho_{2+}(>1/2),\,\rho_{2-}(<1/2).
\end{eqnarray}
These solutions are physically acceptable for $4p/[q(1+p)^2]<1$. 
 For a DW in $T_1$, $\rho_2=\rho_{2-}$. Number conservation then gives
\begin{equation}
 x_{w}=\left(\frac{1+p}{1-p}\right)[n-\rho_{2-}\epsilon_2-\rho_\text{LD}(1-\epsilon_2)].\label{xw11}
\end{equation}
Since a unique solution for $x_w$ can be obtained from (\ref{xw11}), the DW has a fixed position and hence is an LDW. For $x_w>0$ but $<\epsilon_1$, there is an LDW in the bulk of $T_1$, making $\rho_1$ nonuniform, corresponding to the DW-LD-LD phase.
If we set $x_w=0$, all of $T_1,\,T_2$ and $T_3$ are in their LD phases, and is in fact the boundary between the LD-LD-LD phase and DW-LD-LD phase. We get
\begin{equation}
\frac{p}{1+p}(1-\epsilon_2) +\frac{\epsilon_2}{2}\left[1-\sqrt{1-\frac{4p}{q(1+p)^2}}\;\right] =n_{cL}^p,\label{boun1}
\end{equation}
a lower threshold on $n$, such that as $n$ exceeds $n_{cL}^p$ a DW due to PD forms in $T_1$ with the disappearance of the LD-LD-LD phase. As $n$ rises, the LDW shifts first to $T_2$ and then to $T_3$. The corresponding locations can be found by using the logic outlined above. In particular, for an LDW in $T_3$
\begin{equation}
 x_w=\left(\frac{1+p}{1-p}\right)[n+{\epsilon_2}\rho_\text{HD}-\rho_\text{LD}-\frac{\epsilon_2}{2}\rho_{2+}].\label{xw33}
\end{equation}
An LDW can be formed in $T_2$ also; {see Appendix~\ref{ldw-point} for the corresponding LDW position}. In fact, an LDW in $T_2$ assumes a staircase-like shape. 
See Fig.~\ref{pd-ldw-all} {(left)} for a plot of an LDW in $T_2$.
At $x_w=1$, the LDW ends, giving the boundary between the HD-HD-DW and HD-HD-HD phases. We find
\begin{eqnarray}
 \frac{\epsilon_1}{1+p}+ \frac{\epsilon_2}{2}\left[1+\sqrt{1-\frac{4p}{q(1+p)^2}}\right]+\frac{1-\epsilon_1-\epsilon_2}{1+p}=n_{cU}^p,\label{boun2}
\end{eqnarray}
an upper threshold on $n$, such that as $n$ exceeds $n_{cU}^p$, HD-HD-HD phase appears.   


\begin{figure}[htb]
  \includegraphics[height=3.2cm]{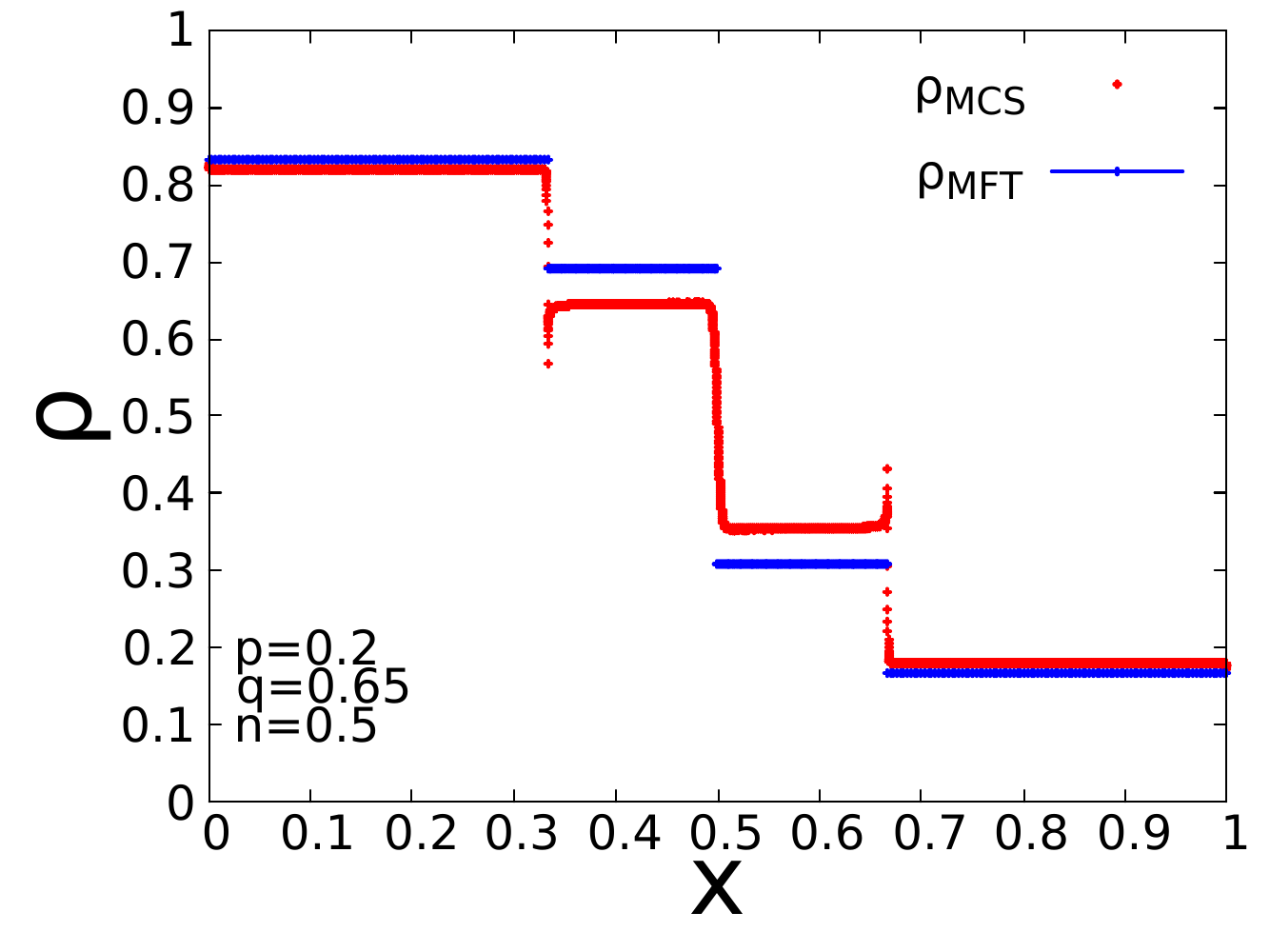} \includegraphics[height=3.2cm]{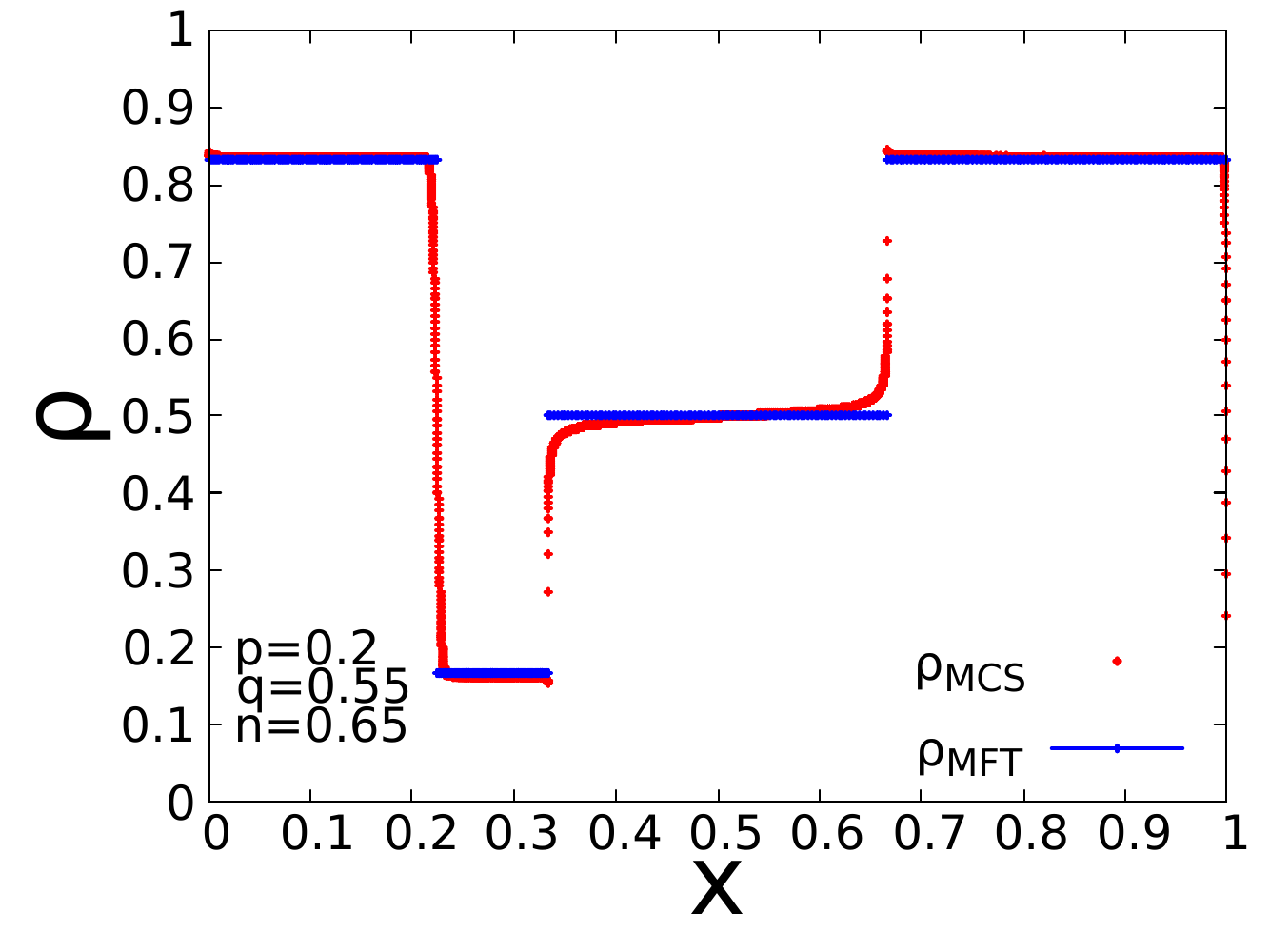}
 \caption{Plots of LDW due to {(left)}  PD  in $T_2$ with $p=0.2,n=0.5,q=0.65,L=9000,\epsilon_1=\epsilon_2=1/3$, {(right)}  ED in $T_1$ with $p=0.2,n=0.65,q=0.55,L=9000,\epsilon_1=\epsilon_2=1/3$. MFT (blue lines) and MCS (red points) results are shown (see text).}\label{pd-ldw-all}
\end{figure}


A DW can however form behind ED, once  $n$ exceeds a {\em different} lower threshold  $n_{cL}^q$. At $n=n_{cL}^q$, $T_2$ reaches its MC phase with $\rho_2=1/2$ in the bulk, and $\rho_1=\rho_3<1/2$. As $n$ is further increased beyond $n_{cL}^q$, a DW is formed in the system, just behind the bottleneck. In the ED dominated regime considered here, $T_2$ is the (extended) bottleneck.
This means a DW is first formed at $x=\epsilon_1+\epsilon_2$ in $T_3$ as soon as $n$ reaches $n_{cL}^q$. As $n$ rises further, the DW starts moving along $T_3$, crossing over to $T_1$ at $x=0$, and then finally reaching $x=\epsilon_1$ at an upper density threshold $n_{cU}^q$, when the DW ends. Any further increase in $n$ will push the system to the HD-HD-HD phase. Thus, the DW {\em never enters} $T_2$, which remains in its MC phase. This is a fundamental difference with the LDW formed due to PD.

With $T_2$ in its MC phase, $\rho_2=1/2$ and $J_2=q/4$. Current conservation then yields ($\rho$ being $\rho_1$ or $\rho_3$)
\begin{equation}
 \rho=\frac{1}{2}\left[1\pm \sqrt{1-{q}}\right]=\rho_+(>1/2),\rho_-(<1/2),\label{rho-ed-sol}
\end{equation}
giving the densities of the high density and low density parts of the DW, which meet at $x_w$.  Using the logic outlined above for a PD together with (\ref{rho-ed-sol}) and particle number conservation, we find with $x_w=\epsilon_1+\epsilon_2$
\begin{equation}
 (1-\epsilon_2)\frac{1}{2}\left[1- \sqrt{1-{q}}\right]+\frac{\epsilon_2}{2}=n\equiv n_{cL}^q,\label{boun3}
\end{equation}
 a critical density above which an LDW due to ED appears that sets the beginning of LD-MC-DW phase. Likewise, assuming an LDW in $T_1$ and setting $x_w=\epsilon_1$ gives
\begin{equation}
 (1-\epsilon_2)\frac{1}{2}\left[1+ \sqrt{1-{q}}\right]+\frac{\epsilon_2}{2}=n\equiv n_{cU}^q,\label{boun4}
\end{equation}
defining an upper threshold on $n$, such that for $n>n_{cU}^q$, HD-HD-HD phase ensues. Our MFT and MCS results on LDW in  $T_1$ is shown in Fig.~\ref{pd-ldw-all} {(right)}. Following the logic outlined above the LDW position in $T_1$ or $T_3$ due to ED can be found straightforwardly; {see Appendix~\ref{ldw-extnd}}.

A minimum current principle determines which of the two possible routes to DW formations is actually realised for a given $(p,q)$.  
The stationary current $J_\text{PD}$ corresponding to a domain wall solution due to PD is  $J_\text{PD}={p}/{(1+p)^2}$. 
 Similarly, the stationary current $J_\text{ED}$ corresponding to a domain wall due to ED is
$ J_\text{ED}=\frac{q}{4}$.
Minimum current principle stipulates that when
$ J_\text{PD}<(>)J_\text{ED}$,
the nonuniform steady states are controlled by the PD (ED), which is consistent with our MCS results. When $J_\text{PD}=J_\text{ED}$, the two defects compete, as we illustrate below, a new kind of states - a pair of delocalised domain walls (DDW) emerges for $n_{cL}^q,\,n_{cL}^p<n < n_{cU}^q,\,n_{cU}^p$. 
Following the logic   of LDW formation in PD and ED dominated regimes, we then expect one DW,  say at $x=x_w^\text{ED}$, due to ED, and another, $x=x_w^\text{PD}$, due to PD. Thus, a complete description of the stationary density profiles require enumeration of both $x_w^\text{PD}$ and $x_w^\text{ED}$. However, with just one condition, {\em viz.}, particle number conservation,  both $x_w^\text{PD}$ and $x_w^\text{ED}$ cannot be uniquely determined. Rather a linear relation between the two is obtained. This means any pair of ($x_w^\text{PD}$, $x_w^\text{ED}$) satisfying particle number conservation is a valid solution. 
Since a DW due to ED must be confined to $T_1$ and $T_3$ only (see above), other DW due to PD must also be confined to the same (even though an LDW due to an isolated PD with $J_\text{PD}<J_\text{ED}$ can also be in $T_2$). Thus, each of ($x_w^\text{PD}$, $x_w^\text{ED}$) must be confined to $T_1$ or $T_3$ only. The inherent stochasticity of the dynamics implies all such ($x_w^\text{PD}$, $x_w^\text{ED}$) pairs satisfying particle number conservation  are visited over time. This further means long-time averages of the densities are {\em inclined} straight lines.  
Both the DWs must have the same height $(1-p)/(1+p)$, with $\rho_\text{HD}=1/(1+p),\,\rho_\text{LD}=p/(1+p),\,\rho_2=1/2$. Particle number conservation gives 
\begin{eqnarray}
&&\rho_\text{HD} x_w^\text{PD}+\rho_\text{LD} (\epsilon_1-x_w^\text{PD})+\frac{\epsilon_2}{2} + \frac{x_w^\text{ED}-\epsilon_1-\epsilon_2}{1+p}\nonumber \\ &&+ \frac{p(1-x_w^\text{ED})}{1+p}=n,\label{ddw-positions}
\end{eqnarray}
 giving a linear relation connecting $x_w^\text{PD}$ and $x_w^\text{ED}$, and not each of the positions separately. As a result, a pair of DDWs is observed. MFT cannot predict the profiles of the DDWs, as it neglects fluctuations. However, we can employ arguments based on symmetry to construct the DDW profiles. With $\epsilon_1=1/3=\epsilon_2$ and exploiting the statistical equivalence of the configurations of the long time averaged envelope of the DDWs in $T_1$ and $T_3$, we 
hypothetically replace each of them by an LDW of height $(1-p)/(1+p)$, connecting  $\rho_\text{HD}=1/(1+p)$ and $\rho_\text{LD}=p/(1+p)$. Such  replacements evidently satisfy particle number conservation. Symmetry of the problem dictates that if $x_0$ is the position of the LDW in $T_1$, the corresponding LDW in $T_3$ must be located at $x_0+2/3$. Application of particle number conservation 
gives
\begin{equation}
 2x_0 \frac{1-p}{1+p}=n-\frac{1}{6}- \frac{2}{3}\left(\frac{p}{1+p}\right).
 \end{equation}
Clearly, if $n=1/2$, $x_0=1/6$, which means the midpoint of the LDW is at the midpoint of $T_1$ or $T_3$, corresponding to DDWs covering entire $T_1$ and $T_3$. If $n<1/2$, $x_0<1/6$, whereas $n>1/2$, $x_0>1/6$, both of which correspond to DDWs partially covering $T_1$ and $T_3$. With the knowledge of $x_0$, it is now possible to obtain the DDW envelope. For instance with $n<1/2$, there is an LD segment in the DDWs in both $T_1$ and $T_3$, which are of length $1/3 - 2x_0$. In other words, the DDW in $T_1$ ($T_3$) wanders a distance $2x_0$, starting from $x=0$ ($x=\epsilon_1+\epsilon_2$). Then joining the densities at $x=0,2x_0$ ($x=\epsilon_1+\epsilon_2,\epsilon_1+\epsilon_2+2x_0$) gives the profiles of the DDW in $T_1$ ($T_2$).
Our MCS results on DDWs in $T_1,\,T_3$ and the corresponding kymographs are shown in Fig.~\ref{ddw-all} {(top, left) and (top, right)} and {Fig.~\ref{ddw-all} (bottom, left) and (bottom, right), respectively}. The analytically obtained DDW profiles are also shown. The kymographs in {Fig.~\ref{ddw-all} (bottom, left) and Fig.~\ref{ddw-all} (bottom, right)} clearly show the synchronised nature of the DDW movements fully or partially covering $T_1$ and $T_3$. See \cite{sm} for a related movie (Movie 4) that visually presents this picture, showing a pair of DDWs, partial or full as controlled by $n$.


 \begin{figure}[htb]
 \includegraphics[height=3.1cm]{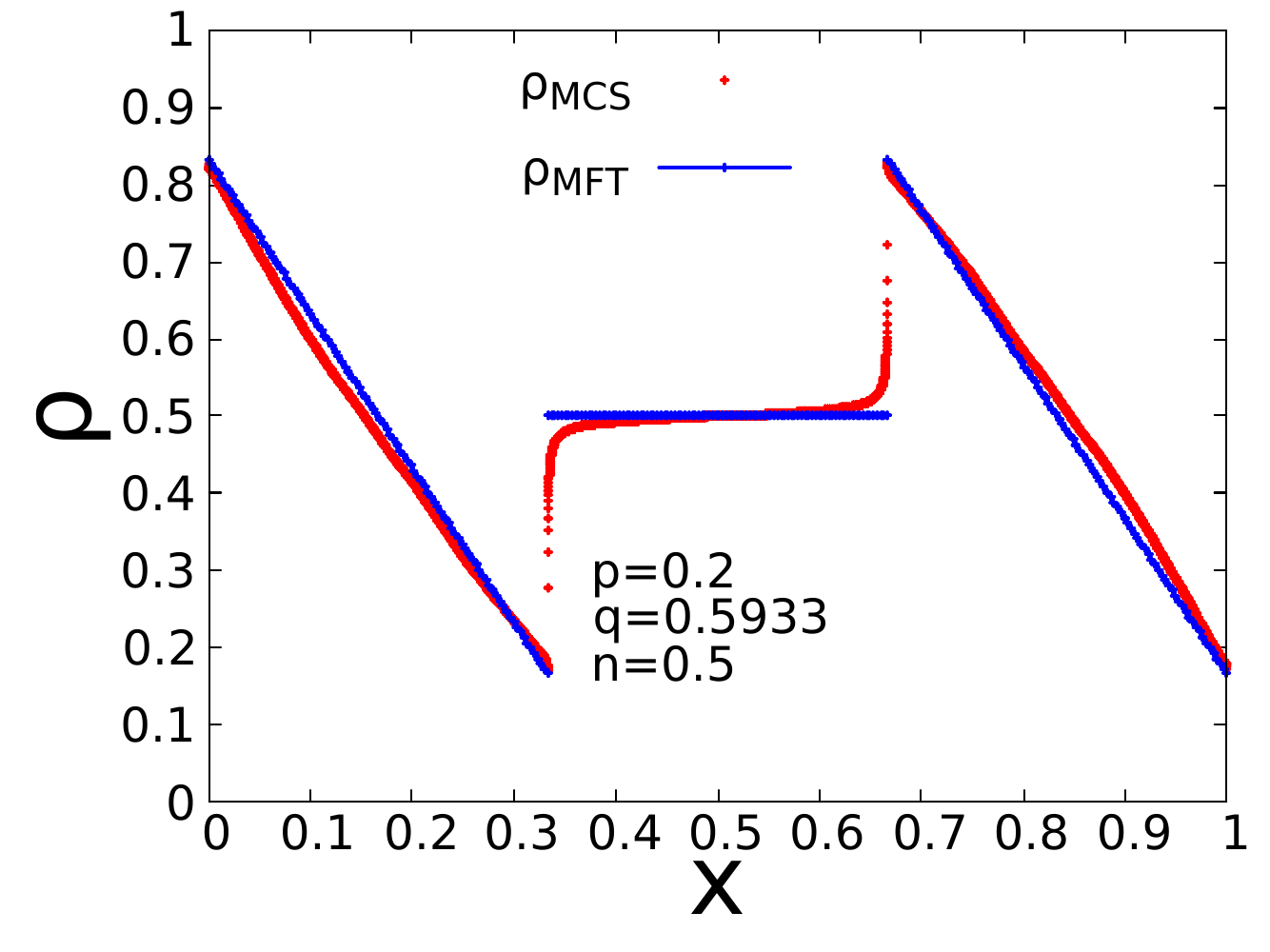}\hfill \includegraphics[height=3.1cm]{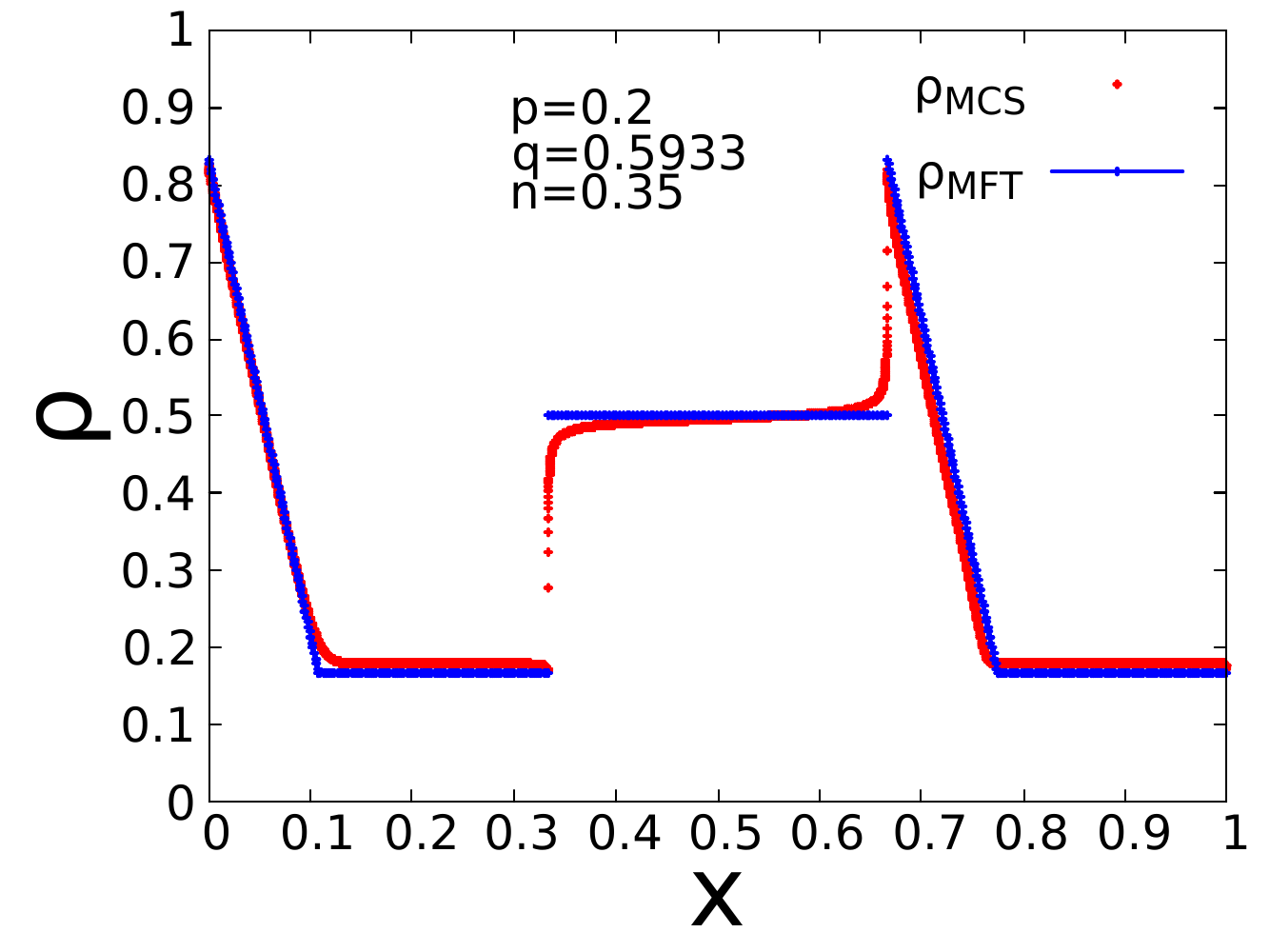}\\
  \includegraphics[width=4.3cm,height=3.3cm]{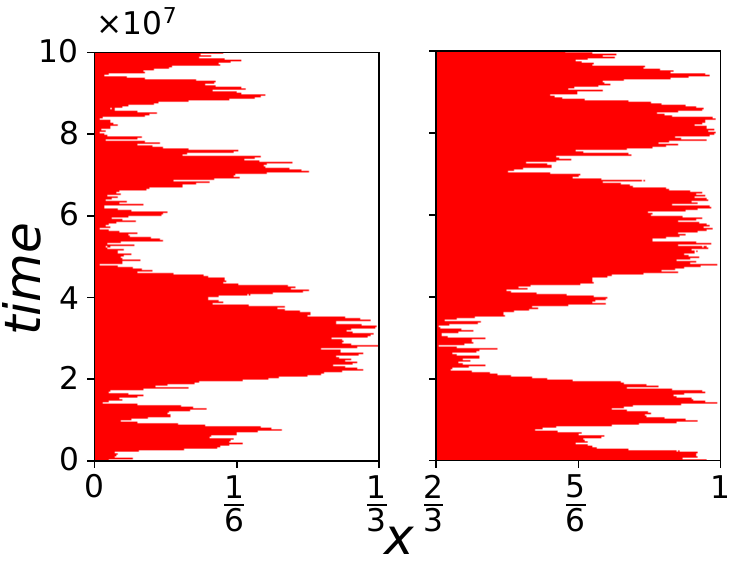}\hfill
  \includegraphics[width=4.3cm,height=3.3cm]{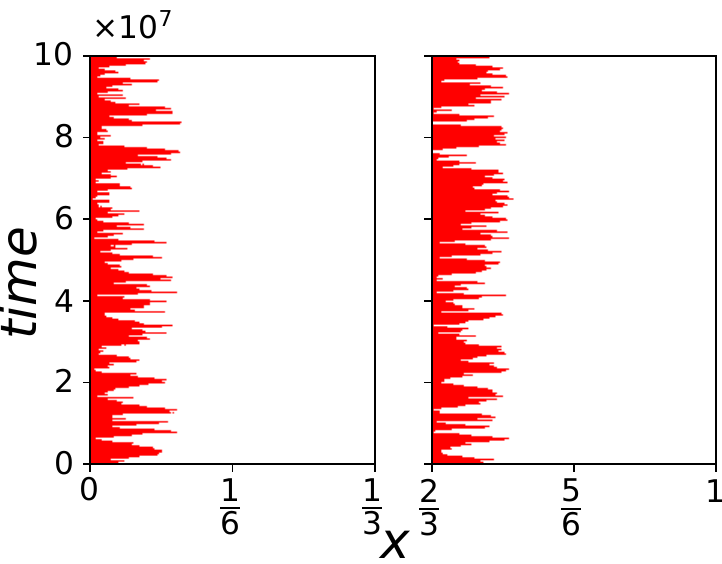}
 \caption{Plots of a pair of DDWs in $T_1$ and $T_3$ with parameters {(top, left)}  $n=0.5,\,p=0.2,\,q=0.5933,\,L=9000,\,\epsilon_1=\epsilon_2=1/3$, {(top, right)} $n=0.35,\,p=0.2,\,q=0.5933,\,L=9000,\,\epsilon_1=\epsilon_2=1/3$, and corresponding kymographs {(bottom, left)} and {(bottom, right)}.
 The analytically obtained DDW profiles (blue lines) and MCS results (red points) are shown (see text).}\label{ddw-all}
\end{figure}


{ Since a DW due to PD {\em can} be inside $T_2$, whereas a DW due to ED {\em cannot} be, an intriguing situation can arise when an LDW due to PD is formed in $T_2$, with $J_\text{PD}<J_\text{ED}$ {\em only slightly}. By tuning $(p,q)$, one can make    $J_\text{PD}>J_\text{ED}$ again {\em only slightly}, for which an LDW is formed due to ED, which however cannot be in $T_2$! Thus the LDW position can jump {\em discontinuously} under a small change in the defect strengths; see movies 5 and 6 in SM~\cite{sm}. As an example, consider $n=1/2$ and $p_c,\,q_c$ satisfy $J_\text{PD}=J_\text{ED}$ together with $\epsilon_1=\epsilon_2=1/3$. Then if $p=p_c-\delta_1,\,q=q_c+\delta_2,\,\delta_1,\,\delta_2>0$ (but $\delta_1,\,\delta_2$ are very small) $J_\text{PD}<J_\text{ED}$, we have an LDW at $x_w=1/2$ due to the point defect. On the other hand, if $\delta_1,\,\delta_2<0$, $J_\text{ED}<J_\text{PD}$ and hence an LDW at $x_w=0$. Thus, as one moves ${\cal O}(\delta),\,\delta=\delta_1,\,\delta_2$ being vanishingly small, the DW position changes by an ${\cal O}(1)$ amount - from $x_w=1/2$ to $x_w=0$. Thus $x_w$ as a function of $\delta_1,\delta_2$ (or $p,q$) shows a jump across the line $J_\text{PD}=J_\text{LD}$: let $\delta \equiv 2\sqrt{\delta_1^2 + \delta_2^2}$ be the ``distance'' between the two points $(p=p_c-\delta_1,\,q=q_c+\delta_2)$ and $(p=p_c+\delta_1,\,q=q_c-\delta_2)$ in the $p-q$ plane. Then, $dx_w/d\delta$ diverges for small $\delta$. Such divergences, a hallmark of defect competition, can occur only when the two points  in the $p-q$ plane lie on the two sides of the line determined by the condition $J_\text{PD}=J_\text{ED}$. This jump in $x_w$ across the line $J_\text{PD}=J_\text{ED}$  indicates a hitherto unknown discontinuous nonequilibrium transition between the PD and ED dominated steady states each with one LDW, with $x_w$ appearing as an order parameter.}

 Phase boundaries (\ref{boun1}), (\ref{boun2}), (\ref{boun3}) and (\ref{boun4}) give the phase diagram in Fig.~(\ref{3d_phase_diagram})  $\epsilon_1=\epsilon_2=1/3$. {For other values of $\epsilon_1,\epsilon_2$, the  phase boundaries change mildly, keeping the topology of the phase diagram unchanged; {see Appendix~\ref{phase-diag-extra} for phase diagrams with (i) $\epsilon_1=2/5,\epsilon_2=1/5$ and (ii) $\epsilon_1=1/4, \epsilon_2=1/2$.}} 

\section{Summary and outlook} 
 
We have thus developed a theory for bottleneck competition in closed, driven single file motion by studying a conceptual model consisting of a periodic TASEP with a point and an extended defects. We show how a dominant defect enforces a particular form of LDW, when sufficient number of particles are available. For competing defects, a pair of DDWs are obtained, instead of a single LDW. The competition between the defects manifests strikingly in the discontinuous jump of the LDW location,  as the system passes from being controlled by one defect to another. Our theory can be generalised to arbitrary number of point or extended defects straightforwardly. { Our MCS studies reveal some finite-size effects and also quantitative disagreement between the MFT and MCS predictions for moderate densities  in the point defect dominated phase space region [see Fig.~\ref{funda} { (left)} and Fig.~\ref{pd-ldw-all} {(left)}]. This can be systematically studied in future by using the finite-size scaling analysis developed in Ref.~\cite{soh}. In contrast, no such disagreements are found in the extended defect dominate phase space regions. These contrasting behaviours may be understood heuristically in terms of an ED working as a particle reservoir for the remaining TASEP, weakening the effects of number conservation; see Ref.~\cite{tirtha-anjan} in a partly related model; see also Ref.~\cite{tirtha-anjan-2} for similar issues.

Our results can be experimentally studied in model experiments
on the collective motion of driven particles with light-induced activity~\cite{light} through a closed narrow circular 
channel~\cite{channel1,channel2}. Suppression of rotational diffusion, e.g., by choosing ellipsoidal particles with the channel width shorter
than the long axis of the particle everywhere, or by using dimer particles can enforce unidirectional movements. Lastly, in spite of the simplicity of our model, the results from it should give insight about the effects of bottlenecks on directional motion in more complex living~\cite{ped1,ped2,ped3} or {\em in-vitro}~\cite{coll1,coll2,coll3} systems. 

{\em Acknowledgement:-} S.M. thanks SERB (DST), India for partial financial support through the CRG scheme [file:
CRG/2021/001875].

\appendix

\section{Mean-field theory for the stationary density profiles}\label{all-ld-hd}


In this Section, we set up and analyze the MFT equations~\cite{blythe} and use them to obtain the LD-LD-LD and HD-HD-HD phase densities.  Let $n_i$ be the occupation at site $i$. In the MFT approximation, correlations are neglected and averages of products are replaced by products of averages~\cite{blythe}. While MFT is an adhoc approximation, it provides a good analytically tractable guideline to the steady state densities and phase diagrams. It is convenient to consider the system to be composed of three segments - $T_1$ between $i=1$ to $\epsilon_1L$, $T_{2}$ between $\epsilon_1L$ to $(\epsilon_1+\epsilon_2)L$  and $T_{3}$ between $(\epsilon_1+\epsilon_2)L$ to $L$, connected serially. The MFT equation for the density $n_i$ in the different segments reads
\begin{eqnarray}
\frac{dn_i}{dt}&=&n_{i+1}[1-n_i]- n_i[1-n_{i-1}],\; 1<i<\epsilon_1L,\\
&=& qn_{i+1}[1-n_i]- qn_i[1-n_{i-1}],\nonumber \\
&&\;\;\;\;\;\;\epsilon_1L\leq i \leq (\epsilon_1+\epsilon_2)L,\\
&=& n_{i+1}[1-n_i]- n_i[1-n_{i-1}],\nonumber\\
&&\;\;\;\;\;\; (\epsilon_1+\epsilon_2)L< i\leq L.
 \end{eqnarray}
 We note that the above mean-field equations are invariant under the transformations $n_i\rightarrow 1-n_{L-i-1}$, which defines the particle hole symmetry in this model.
Before proceeding further it is convenient to introduce a quasi-continuous coordinate $x\equiv i/L$ in the thermodynamic limit $L\rightarrow \infty$. Thus, we have $0\leq x\leq 1$. Now define $\rho_a(x)\equiv \langle n^a_i\rangle$, where $\langle...\rangle$ implies temporal averages in the steady states and $a=1,\,2$ or $3$ for the three segments.

We recall, as given the main text, that the stationary currents $J_1,\,J_2,\,J_3$ in the three segments in MFT 
\begin{eqnarray}
 J_1&=&\rho_1(1-\rho_1),
 \label{curr1-app}\\
 J_{2}&=&q\rho_2(1-\rho_2),\label{curr2}\\
 J_{3}&=& \rho_3(1-\rho_3) \label{curr3},
\end{eqnarray}
where $\rho_1,\,\rho_2$ and $\rho_3$ are the stationary densities in $T_1,\,T_{2}$ and $T_{3}$ respectively, are all equal due to current conservation:
\begin{equation}
J_1=J_{2}=J_{3}\label{curr4}
\end{equation}
in the steady states.

\subsection{LD-LD-LD phase}

For sufficiently  low mean densities $n$, all the channels should be sparsely populated, and hence we expect all the three segments to be in their LD phases with uniform densities $\rho_1=\rho_3<\rho_2<1/2$ respectively. These may be obtained as follows. By using  $J_1=J_{2}=J_{3}$
\begin{equation}
 \rho_1(1-\rho_1)=q\rho_2(1-\rho_2)=\rho_3(1-\rho_3).\label{curr-cons}
 \end{equation}
Now using particle number conservation,  we get
\begin{equation}
 [1-\epsilon_2]\rho_1 +\rho_2\epsilon_2=n. \label{pnc-ld}
\end{equation}
For explicit solutions of the densities,   (\ref{curr-cons}) together with (\ref{pnc-ld}) can be used to give
\begin{equation}
 \rho_1=\left[-B\pm \sqrt{B^2 -4AC}\right]\frac{1}{2A},\label{rho1-sol}
\end{equation}
as the two general solutions, where
\begin{eqnarray}
 A&=&(1-\epsilon_2)^2 - \epsilon_2^2/q,\\
 B&=&\frac{\epsilon_2^2}{q}-(1-\epsilon_2)(2n-\epsilon_2),\\
 C&=& n^2 -\epsilon_2 n.
\end{eqnarray}
So far we have not imposed any conditions of the LD-LD-LD phase on the solutions (\ref{rho1-sol}). The pertinent question then is:
Which of the two solutions in (\ref{rho1-sol}) is to be considered as the LD-LD-LD phase density solution? To settle this, we use the fact that in the limiting case with vanishing particle number, i.e., with $n\rightarrow 0$, the LD-LD-LD phase solution must smoothly go to zero. This consideration allows us to pick the right solution in (\ref{rho1-sol}) for the LD-LD-LD phase: we choose the solution that vanishes as $n\rightarrow 0$. Which one among the two in (\ref{rho1-sol}) does that depends upon the signs of $A$ and $B$, i.e., will be decided by $\epsilon_2$ and $n$. 
We thus note that the point defect has no macroscopic effect on the steady state density profiles. Instead at $x=0$, the location of the point defect, there is a local peak of height $h$ with vanishing width in the thermodynamic limit, where $h=\rho_1(1-p)/p$, such that the local density at $x=0$ is $\rho_1 + h$~\cite{niladri-tasep,erwin-defect}. This essentially acts as a boundary layer between $T_1$ and $T_3$. Steady state density profiles in the LD-LD-LD phase with $n=0.15, p=0.2,\,q=0.45$ and $n=0.15, p=0.15,\,q=0.6$ with a system size $L=9000$ and $\epsilon_1=\epsilon_2=1/3$ are shown in 
Fig.~\ref{ld-plot} {(top) and Fig.~\ref{ld-plot} (bottom)} respectively. Good agreement between MFT and MCS results are observed.
\begin{figure}[htb]
 \includegraphics[width=8.5cm]{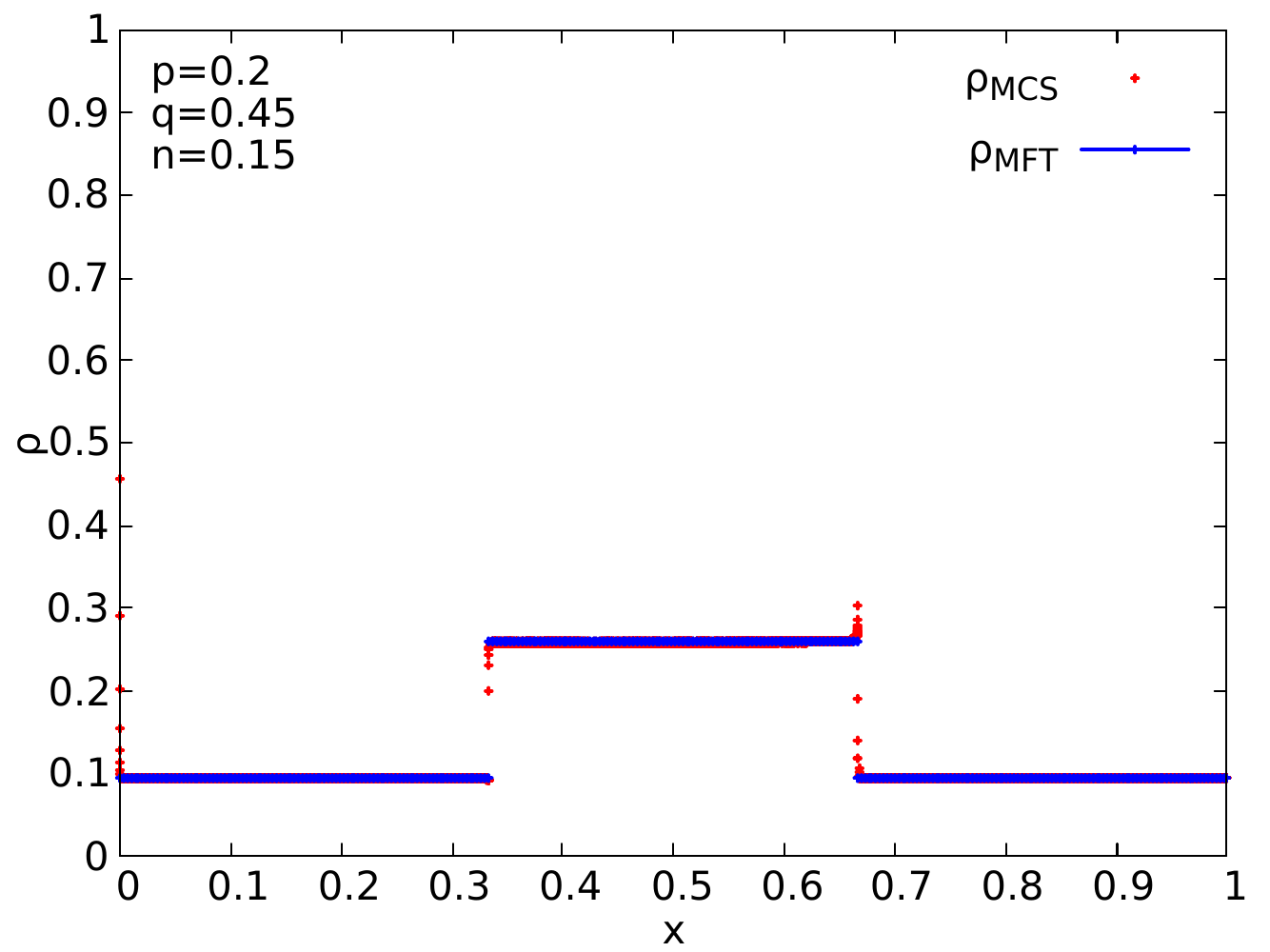}\hfill
 \includegraphics[width=8.5cm]{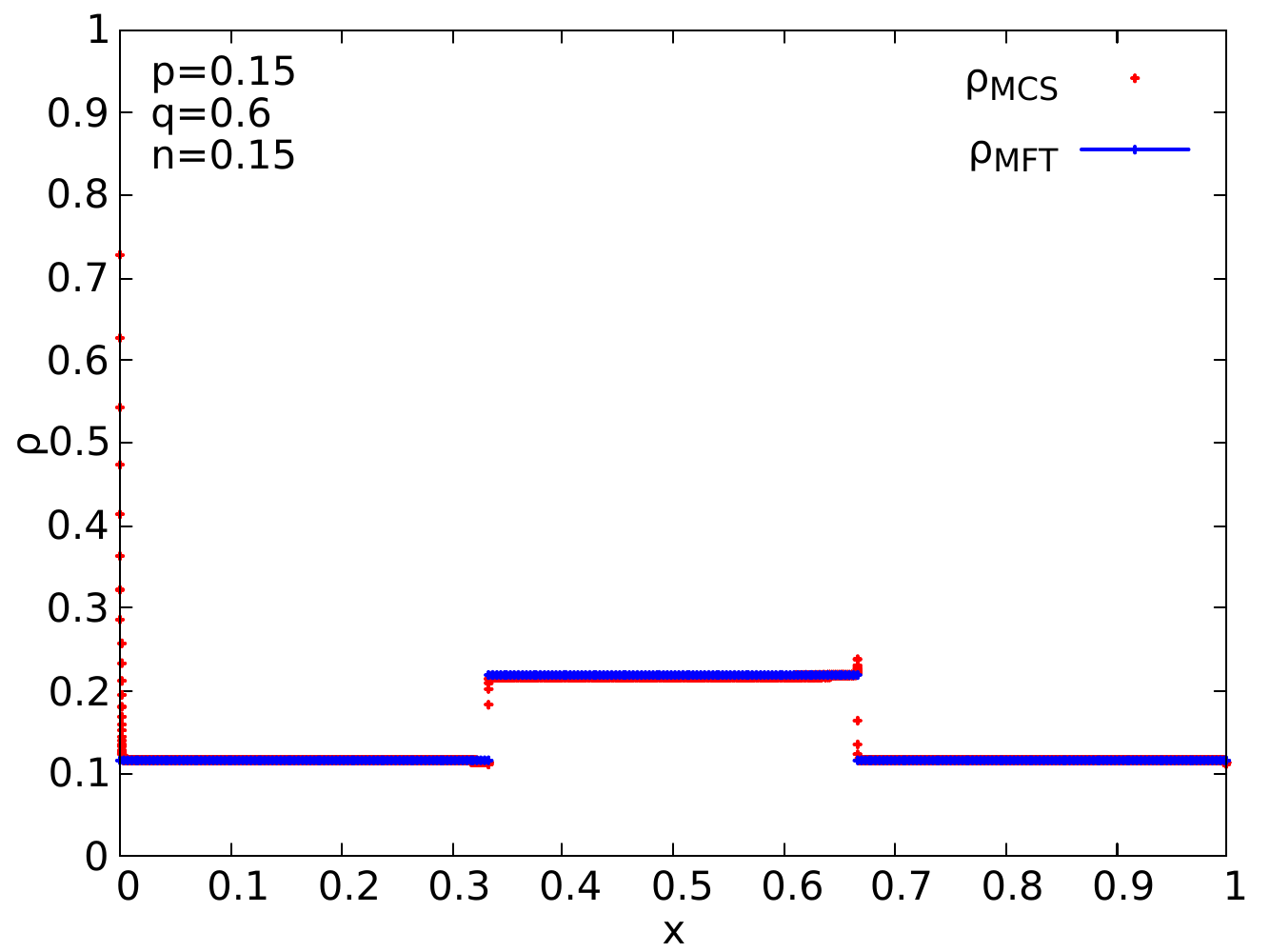}
 \caption{LD-LD-LD density plots with $n=0.15$: {(top)} $p=0.2,\,q=0.45$, {(bottom)} $p=0.15,\,q=0.6$. MFT (blue lines) and MCS (red points) results are shown. Good agreement between MFT and MCS results are observed. }\label{ld-plot}
\end{figure}

\subsection{HD-HD-HD phase}

The HD-HD-HD phase can be analysed by applying the particle-hole symmetry on the LD-LD-LD phase.  Physically, for very high $n$, all of $T_1,T_2$ and $T_3$ should be nearly filled with particles, and hence HD-HD-HD phase is expected. In this case, $\rho_1,\rho_2,\rho_3>1/2$ and $\rho_1=\rho_3>\rho_2$.

For an explicit solution for the HD-HD-HD phase, we again consider (\ref{rho1-sol}) and note that if $n=1$, unsurprisingly $\rho_1=1$ is a solution, which in turn means $\rho_2=1$ and $\rho_3=1$, i.e., a completely filled up system. Thus, when the system is nearly filled and all of $T_1,T_2,T_3$ are in their HD phases, we should accept that particular solution in (\ref{rho1-sol}) which smoothly reaches unity when $n\rightarrow 1$. Which of the two solutions in (\ref{rho1-sol}) will satisfy this property depends on the signs of $A,\,B$.

Similar to the LD-LD-LD phase, the point defect has no macroscopic effect on the stationary densities. Instead, one has a local dip of depth $h'$, with vanishing width in the thermodynamic limit, where $h'=(1-\rho_1)(1-p)/p$, such that at $x=0$, the density is $\rho_1-h'$. Steady state density profiles in the HD-HD-HD phase with $n=0.8, p=0.2,\,q=0.45$ and $n=0.8, p=0.15,\,q=0.6$ with a system size $L=9000$ and $\epsilon_1=\epsilon_2=1/3$ are shown in 
{Fig.~\ref{hd-plot} (top) and Fig.~\ref{hd-plot} (bottom) respectively}. Good agreement between MFT and MCS results are observed.
\begin{figure}[htb]
 \includegraphics[width=8.5cm]{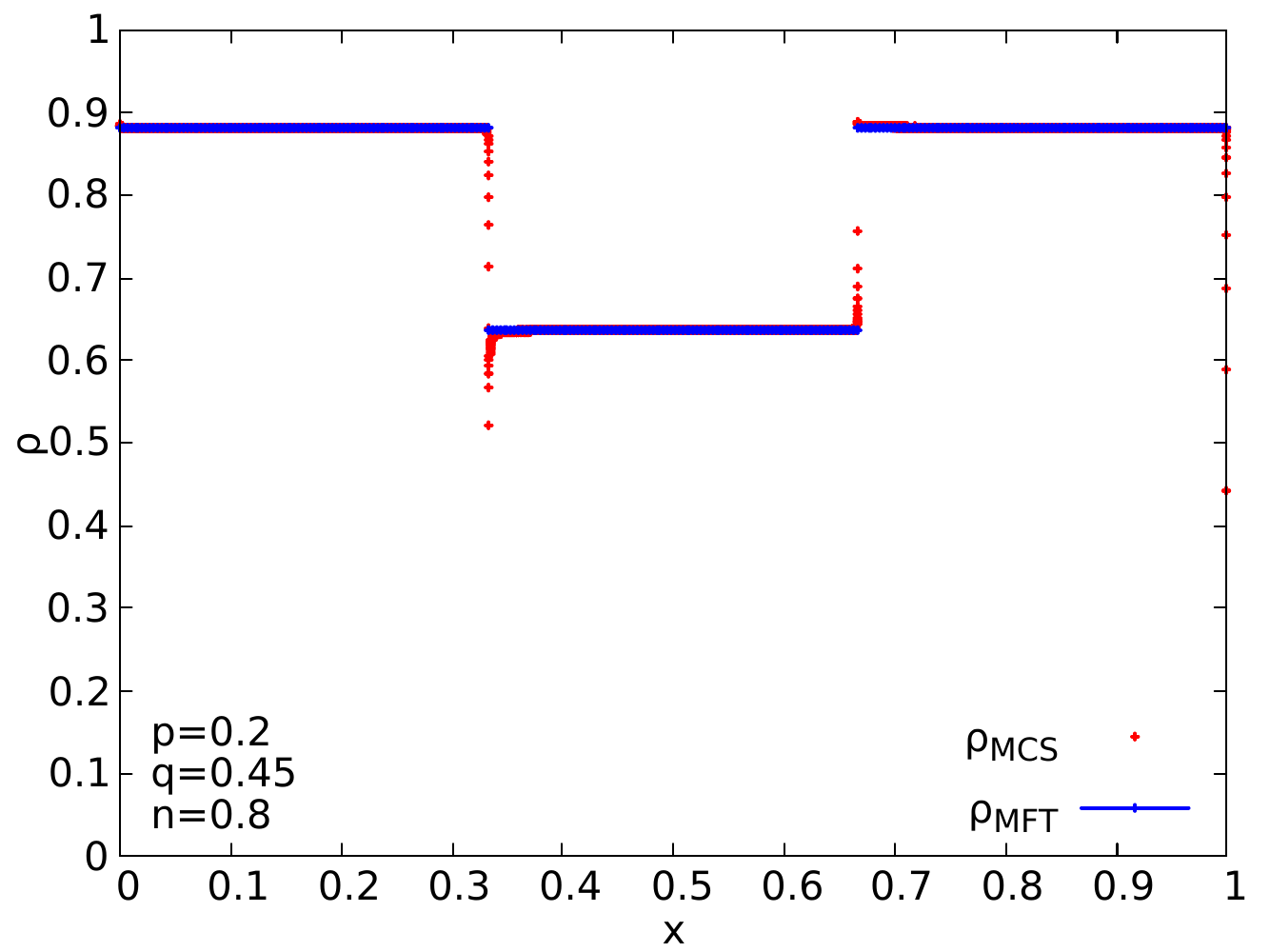}\hfill
 \includegraphics[width=8.5cm]{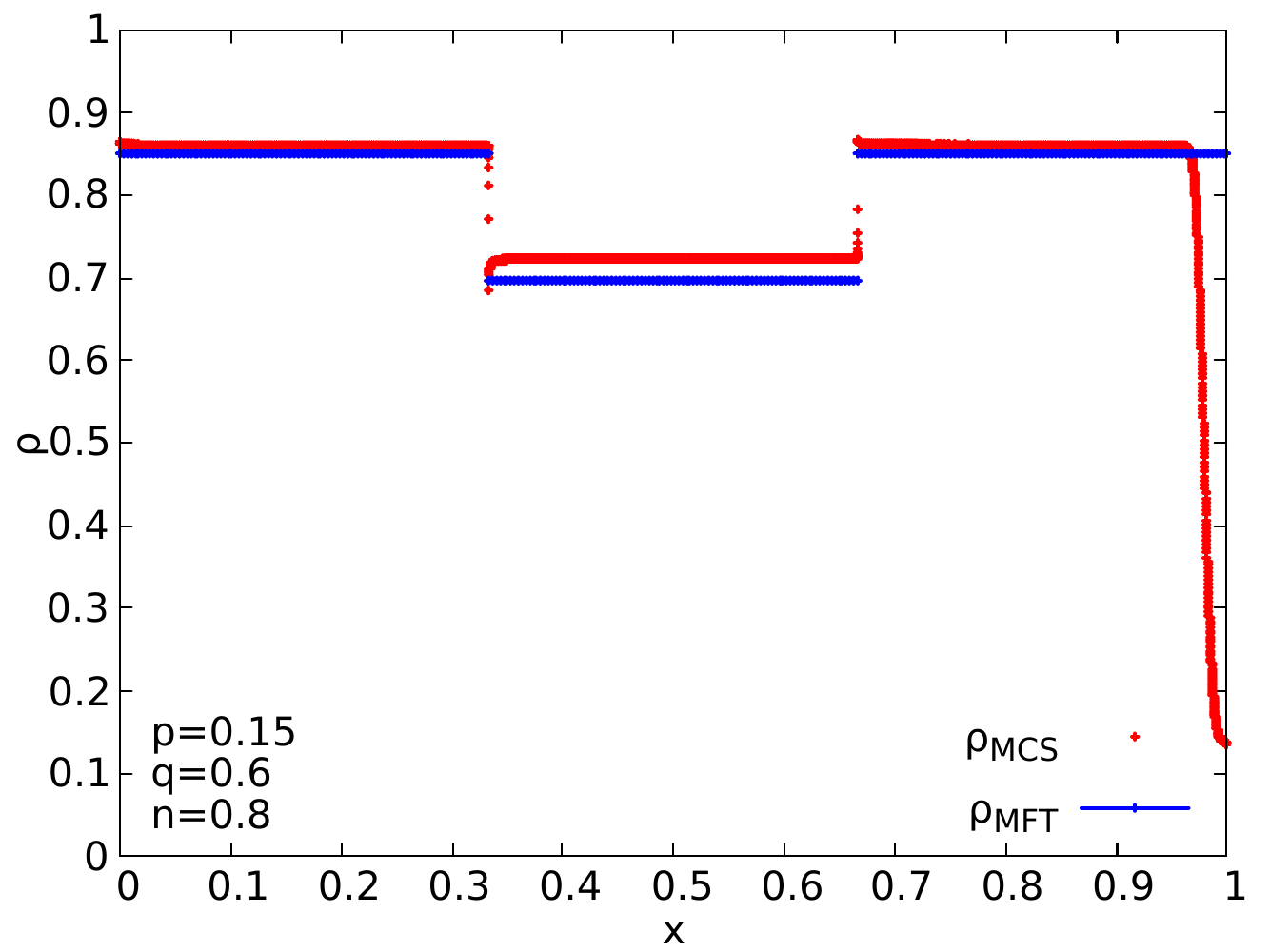}
 \caption{HD-HD-HD density plots: {(top)} $n=0.8,\,p=0.2,\,q=0.45$, {(bottom)} $n=0.8,\,p=0.15,\,q=0.6$. MFT (blue lines) and MCS (red points) results are shown. Good agreement between MFT and MCS results are observed. }\label{hd-plot}
\end{figure}

\subsection{MC-MC-MC phase}

In the MC phase, the density should be 1/2. This means in these putative MC phases, $J_1=J_3=1/4,\,J_2=q/4$. This immediately shows that there is no MC-MC-MC phase in the model, i.e., all the three segments cannot be simultaneously in the MC phases, as that would violate the current conservation condition (\ref{curr-cons}). 

\section{LDWs due to the point defect}\label{ldw-point}

In the main text, we have presented a plot of LDW in $T_2$ due to the point defect, giving a staircase-like stationary density profile, as obtained from our MFT and MCS studies. As mentioned in the main text, an LDW due to a point defect can be {\em anywhere} in the ring, i.e., it can be in $T_1$ and $T_3$ as well. 

First consider the case when the LDW is inside $T_1$: $0\leq x_w\leq \epsilon_1$. Since the particles flow in the anticlockwise direction, and an LDW should form {\em behind} a bottleneck (which in the present case is the point defect at $x=0$), we write
\begin{eqnarray}
 \rho_1(x)&=&\rho_\text{HD},\;\;0\leq x\leq x_w,\nonumber \\
          &=&\rho_\text{LD},\;\;x_w\leq x\leq \epsilon_1,
\end{eqnarray}
In this situation, we must have the entire $T_2$ to be its LD phase: $\rho_2(x)=\rho_{2-}$ for $\epsilon_1\leq x\leq \epsilon_1+\epsilon_2$. In addition, $T_3$ too should be in its LD phase, with $\rho_3=\rho_\text{LD}$. We can now apply PNC to determine $x_w$. We get
 \begin{equation}
  \rho_\text{HD}x_w + \rho_\text{LD}(\epsilon_1-x_w)+\rho_{2-}\epsilon_2 + \rho_\text{LD}(1-\epsilon_1-\epsilon_2)=n.\label{xw1}
 \end{equation}
Equation~(\ref{xw1}) gives $x_w$ as a function of $p,q$ and $n$. Simplifying (\ref{xw1}), we obtain
\begin{eqnarray}
 x_{w}&=&\left(\frac{1+p}{1-p}\right)[n-\rho_{2-}\epsilon_2-\rho_\text{LD}(1-\epsilon_2)],\label{xw11-app}
\end{eqnarray}
giving a unique solution for $x_w$, as it should be for an LDW.
For $x_w>0$ but $<\epsilon_1$, there is an LDW in the bulk of $T_1$, making $\rho_1$ nonuniform, corresponding to the DW-LD-LD phase.

When the LDW is inside $T_2$, $T_1$ is in its HD phase, hence $\rho_1=\rho_\text{HD}$, and $T_3$ is in its LD phase, with $\rho_3=\rho_\text{LD}$. Further, since the LDW is located inside $T_2$, we must have $\rho_2=\rho_{2+}$ for $\epsilon_1<x<x_w$; $\rho_2=\rho_{2-}$ for $x_w<x<\epsilon_1+\epsilon_2$. Applying PNC we get
\begin{widetext}
\begin{equation}
 \rho_\text{HD}\epsilon_1+ (x_w-\epsilon_1)\rho_{2+}+ (\epsilon_1+\epsilon_2-x_w)\rho_{2-}+\rho_\text{LD}(1-\epsilon_1-\epsilon_2)=n. \label{pnct2}
\end{equation}
Solving (\ref{pnct2}), we get
\begin{eqnarray}
 x_w&=& \frac{1}{(\rho_{2+}- \rho_{2-})}\bigg[n+\rho_{2+}\epsilon_1-\rho_\text{HD}\epsilon_1- (\epsilon_1+\epsilon_2)\rho_{2-}-\rho_\text{LD}(1-\epsilon_1-\epsilon_2)\bigg]\label{xw22}
\end{eqnarray}
giving a unique position of the LDW in $T_2$. 
\end{widetext}
The overall density profile, consisting of HD phase in $T_1$ and LD phase in $T_3$, with an intervening LDW in $T_2$ takes the form of a step-like structure. This is specifically attributed to $T_2$ being an extended defect with a hopping rate $q<1$. This is shown in Fig.~3 {(left)} of the main text.

Next, consider  an LDW in $T_3$. Thus, $\rho_3(x)=\rho_\text{HD},\;\epsilon_1+\epsilon_2<x<x_w$, $\rho_3(x)=\rho_\text{LD},\;x_w<x<1$. Further, $\rho_1(x)=\rho_\text{HD}$, $\rho_2(x)=\rho_{2+}$, since both $T_1$ and $T_2$ are in their HD phases. Then applying PNC, we obtain
\begin{widetext}
\begin{eqnarray}
  &&\frac{\epsilon_1}{1+p}+ \frac{\epsilon_2}{2}\left[1+\sqrt{1-\frac{4p}{q(1+p)^2}}\;\right]+(x_w-\epsilon_1-\epsilon_2)\frac{1}{1+p} + \frac{(1-x_w)p}{1+p}=n. \label{pnct3}
\end{eqnarray}
Solving, we find
\begin{equation}
 x_w=\frac{1+p}{1-p}\left[n+\frac{\epsilon_2}{1+p}-\frac{p}{1+p}-\frac{\epsilon_2}{2}\left[1+\sqrt{1-\frac{4p}{q(1+p)^2}}\right]\right]\label{xw33-app}
\end{equation}
as the LDW position in $T_3$.
\end{widetext}

Our MFT and MCS results on LDWs in $T_1,\,T_3$ are shown in { Fig.~\ref{pd-ldw-1-3} (top)
and Fig.~\ref{pd-ldw-1-3} (bottom)} respectively. We find reasonable agreement between our MFT and MCS results.
\begin{figure}[htb]
 \includegraphics[height=6cm]{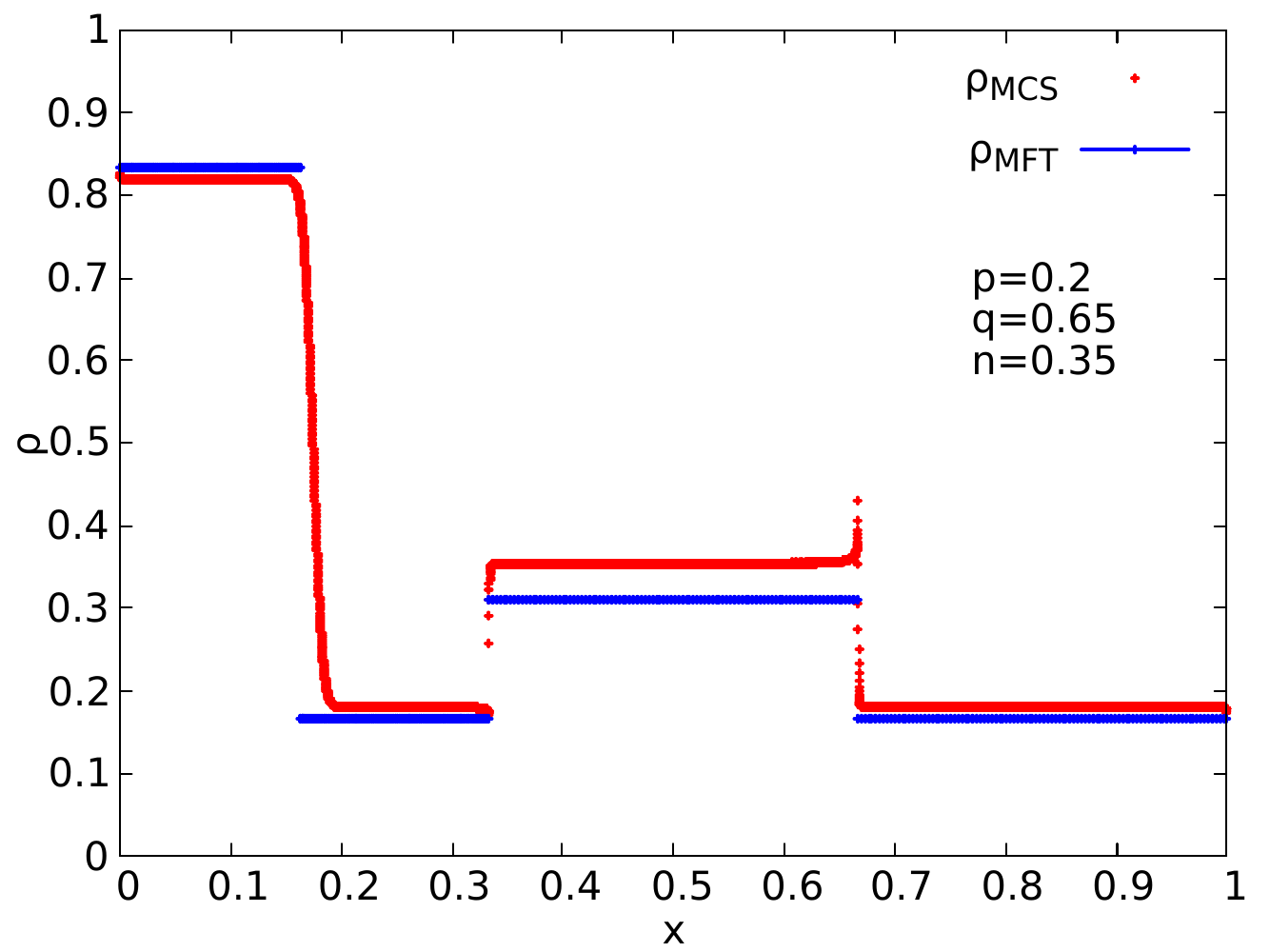}\hfill \includegraphics[height=6cm]{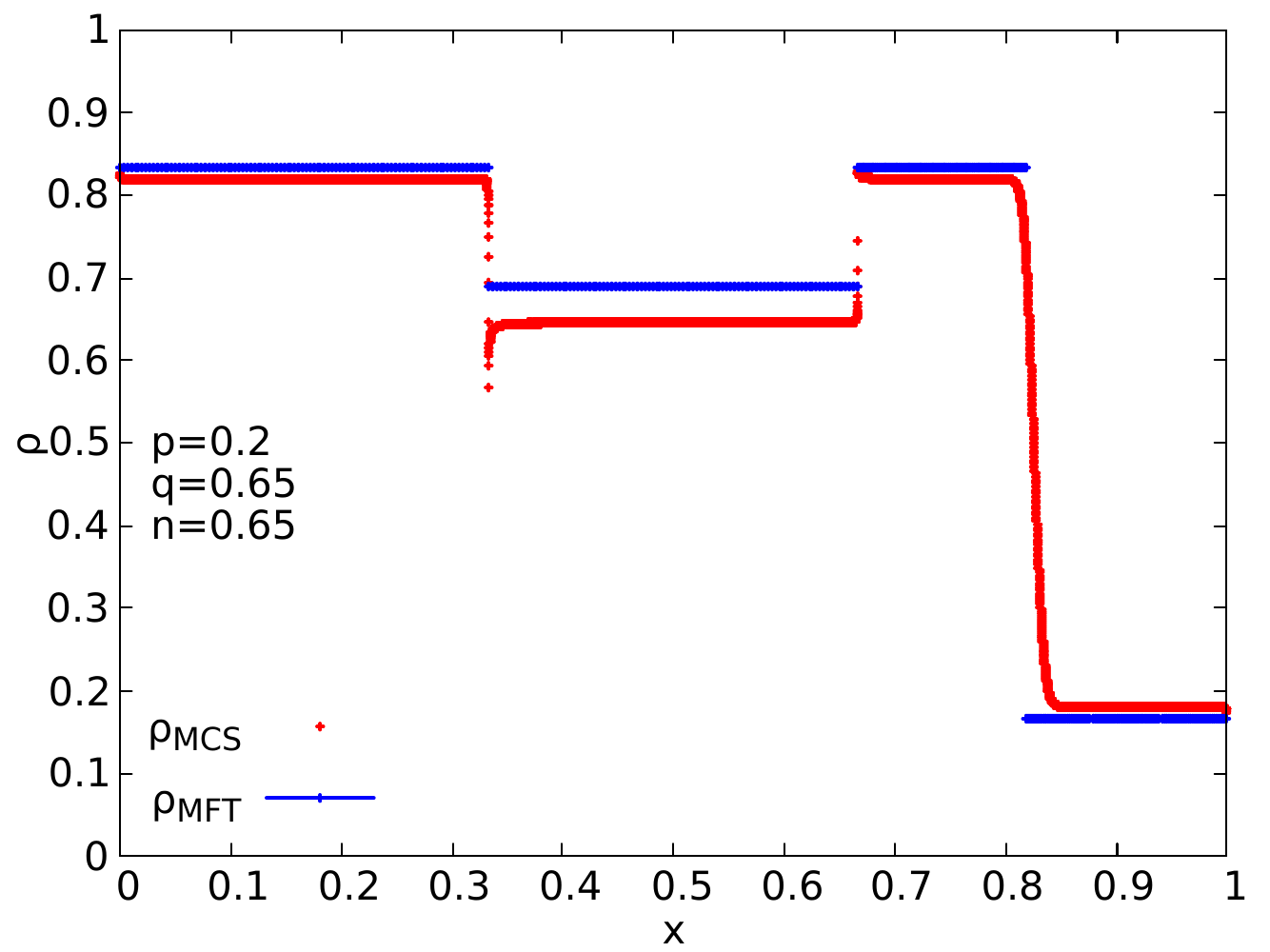}
 \caption{Plot of an LDW in {(top)} $T_1$ in the point defect dominated regime with $p=0.2,n=0.35,q=0.65,L=9000,\epsilon_1=\epsilon_2=1/3$, and {(bottom)} $T_3$ in the point defect dominated regime with $p=0.2,n=0.65,q=0.65,L=9000,\epsilon_1=\epsilon_2=1/3$. MFT (blue lines) and MCS (red points) results are shown (see text).}\label{pd-ldw-1-3}
\end{figure}

\section{LDWs due to the extended defect}\label{ldw-extnd}

As mentioned in the main text, an LDW can be formed due to the extended defect as well. In this case, $T_2$ is in its MC phase with $\rho_2=1/2$, and an LDW can be formed in either $T_1$ or $T_3$. In the main text, we have shown an LDW in $T_1$ due to the extended defect obtained from our MFT and MCS studies. Here, we present an LDW in $T_3$  due to the extended defect in Fig.~\ref{ed-ldw-3}.

Current conservation yields
\begin{equation}
 \rho(1-\rho)=\frac{q}{4},
\end{equation}
where $\rho$ is $\rho_1$ or $\rho_3$. Solving,
\begin{equation}
 \rho=\frac{1}{2}\left[1\pm \sqrt{1-{q}}\right]=\rho_+(>1/2),\rho_-(<1/2),\label{rho-ed-sol-app}
\end{equation}
giving the densities of the high density and low density parts of the DW, which meet at $x_w$. As before, $x_w$ can be calculated by using PNC. Assume an LDW in $T_3$.  This is the LD-MC-DW phase. Then PNC gives
\begin{equation}
 \rho_-\epsilon_1 +\frac{\epsilon_2}{2}+ (x_w-\epsilon_1-\epsilon_2)\rho_++(1-x_w)\rho_-=n.\label{q-dw}
 \end{equation}
 Setting $x_w=\epsilon_1+\epsilon_2$ gives the condition for transition from the LD-LD-LD phase to LD-MC-DW phase, when the DW is due to the extended defect. We get
\begin{equation}
 \epsilon_1\rho_-+\frac{\epsilon_2}{2}+(1-\epsilon_1-\epsilon_2)\rho_-=n.\label{boun4-app}
\end{equation} 
This is independent of $p$, but depends upon $q$ through the dependence of $\rho_-$ on $q$. Using (\ref{rho-ed-sol-app}), we find
\begin{equation}
 (1-\epsilon_2)\frac{1}{2}\left[1- \sqrt{1-{q}}\right]+\frac{\epsilon_2}{2}=n\equiv n_{cL}^q,
\end{equation}
defining a critical density $n_{cL}^q$ above which an LDW due to the extended defect appears.

\begin{figure}[htb]
 \includegraphics[width=8cm]{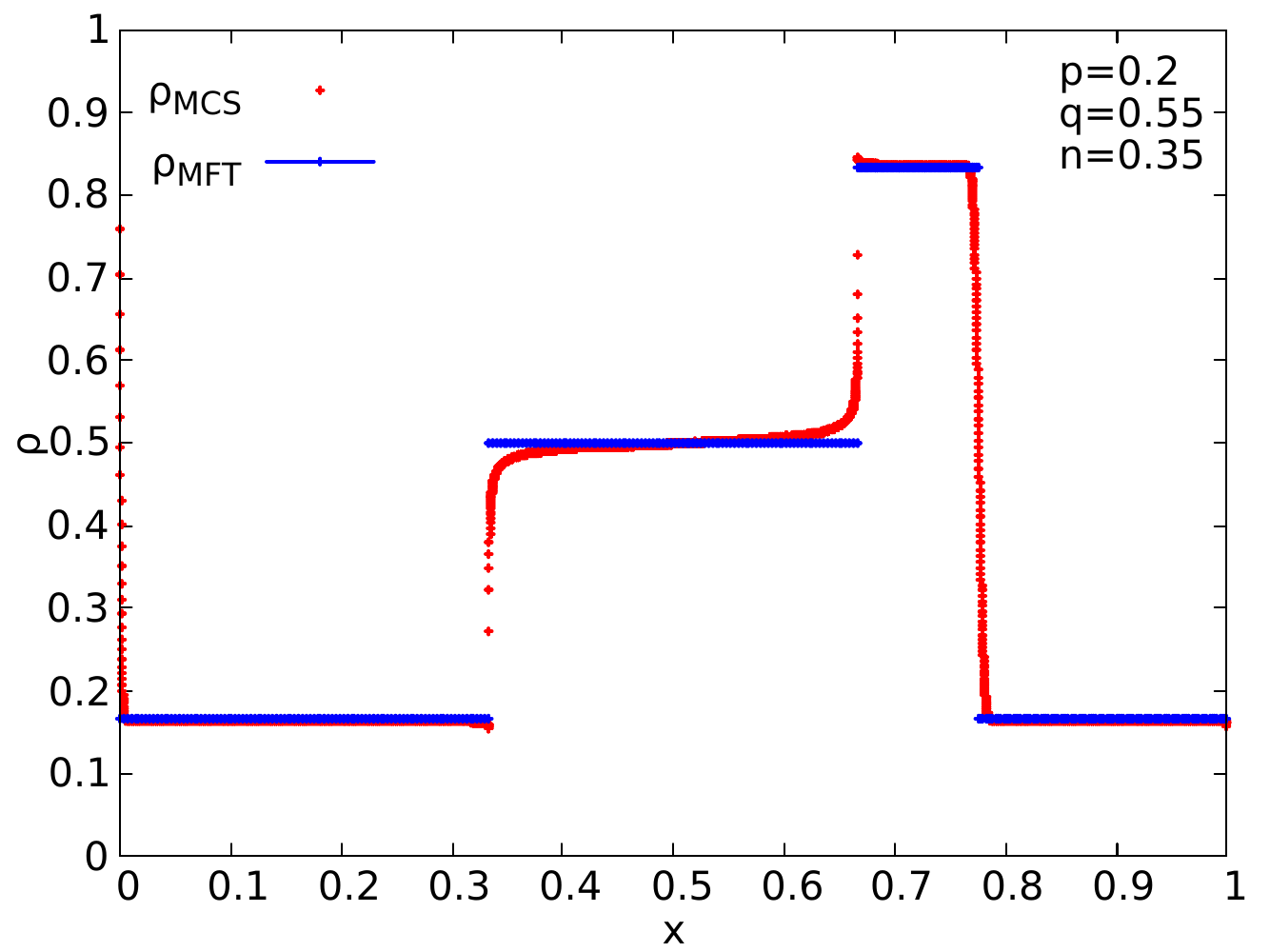}
 \caption{Plot of an LDW in $T_3$ in the extended defect dominated regime with $p=0.2,n=0.35,q=0.55,L=9000,\epsilon_1=\epsilon_2=1/3$. MFT (blue lines) and MCS (red points) results are shown (see text).}\label{ed-ldw-3}
\end{figure}

Now consider an LDW in $T_1$. In this case, $T_3$ is in the HD phase. PNC gives
\begin{equation}
 x_w\rho_{+} + (\epsilon_1-x_w)\rho_{-} + \frac{\epsilon_2}{2}+(1-\epsilon_1-\epsilon_2)\rho_+  =n.
\end{equation}
Setting $x_w=\epsilon_1$ produces the boundary between the DW-MC-HD phase and HD-HD-HD phase:
\begin{equation}
 (1-\epsilon_2)\rho_+ +\frac{\epsilon_2}{2}=n.\label{boun5}
\end{equation}
Substituting for $\rho_+$, we get
\begin{equation}
 (1-\epsilon_2)\frac{1}{2}\left[1+ \sqrt{1-{q}}\right]+\frac{\epsilon_2}{2}=n\equiv n_{cU}^q,
\end{equation}
defining an upper threshold on $n$, such that for $n>n_{cU}^q$, HD-HD-HD phase is predicted.

\section{Phase diagrams}\label{phase-diag-extra}

In this Section, we give the mean-field  phase diagrams for our model with unequal segments: We consider (i) $\epsilon_1=2/5,\epsilon_2=1/5$ and (ii) $\epsilon_1=1/4, \epsilon_2=1/2$; see { Fig.~\ref{phase-sm} (top) and Fig.~\ref{phase-sm} (bottom)} respectively. These two phase diagrams and the phase diagram in the main text (with $\epsilon_1=1/3,\epsilon_2=1/3$) have the same topology, with the phase boundaries shifting mildly. Thus the phase diagrams do not sensitively depend on the sizes of the segments, controlled by $\epsilon_1$ and $\epsilon_2$. 

\begin{figure}[htb]
 \includegraphics[width=7cm]{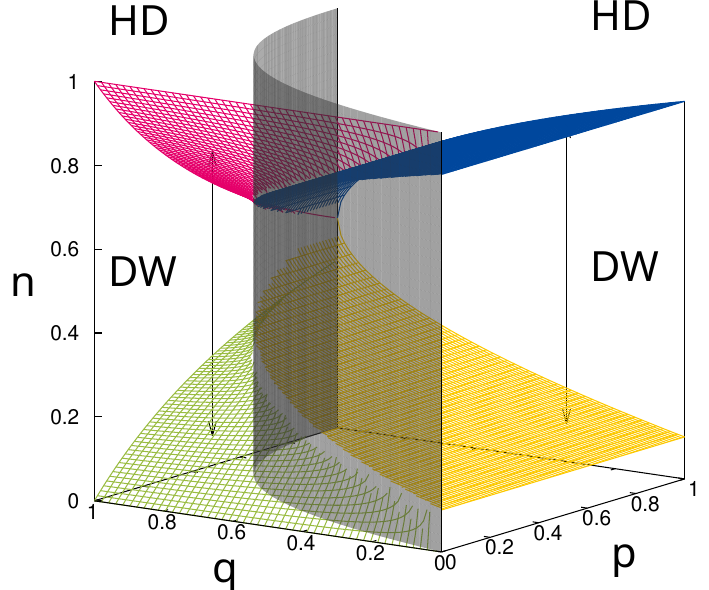}\hfill \includegraphics[width=7cm]{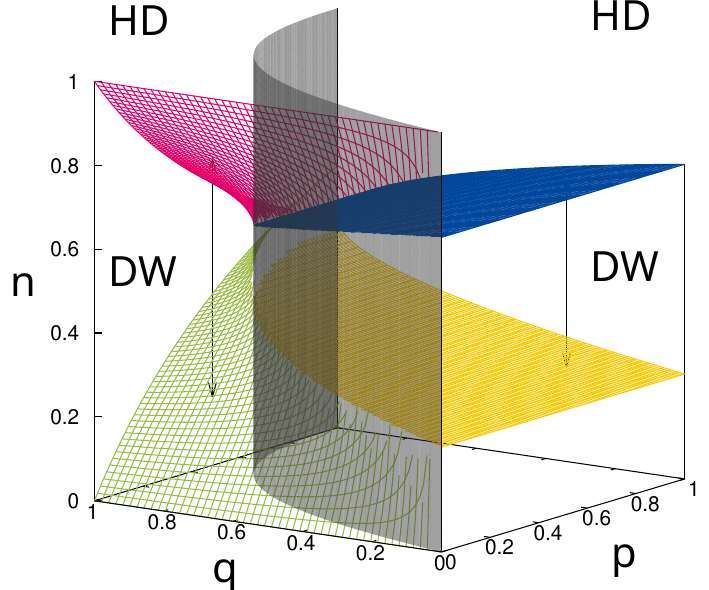}.
 \caption{Mean-field phase diagrams with parameters {(top)}  $\epsilon_1=2/5,\epsilon_2=1/5$ and {(bottom)} $\epsilon_1=1/4, \epsilon_2=1/2$. These have the same topology and are very similar to the phase diagram in the main text (with $\epsilon_1=1/3,\epsilon_2=1/3$)} \label{phase-sm}
\end{figure}

\end{document}